\documentclass[prb,superscriptaddress,showpacs,twocolumn,amsmath,amssymb,floatfix]{revtex4-1}

\usepackage{graphicx}
\usepackage{epsfig}
\usepackage{bm}
\usepackage{color}
\usepackage[applemac]{inputenc}
\usepackage{subfig}
\DeclareGraphicsExtensions{.pdf .eps, .png, .jpg}
\begin{document}
   
\title{Effective mass theory for the anisotropic exciton in 2D crystals: Application to phosphorene}

\author{Elsa Prada}
  \affiliation{Departamento de F\'{\i}sica de la Materia Condensada, Universidad Aut\'onoma de Madrid, Cantoblanco, 28049 Madrid, Spain}
   \affiliation{Instituto de Ciencia de Materiales Nicol\'as Cabrera, Universidad Aut\'onoma de Madrid, Cantoblanco, 28049 Madrid, Spain}
    \affiliation{Condensed Matter Physics Center (IFIMAC),\\ Universidad Aut\'onoma de Madrid, Cantoblanco, 28049 Madrid, Spain}
 \author{J. V. Alvarez}
   \affiliation{Departamento de F\'{\i}sica de la Materia Condensada, Universidad Aut\'onoma de Madrid, Cantoblanco, 28049 Madrid, Spain}
   \affiliation{Instituto de Ciencia de Materiales Nicol\'as Cabrera, Universidad Aut\'onoma de Madrid, Cantoblanco, 28049 Madrid, Spain}
    \affiliation{Condensed Matter Physics Center (IFIMAC),\\ Universidad Aut\'onoma de Madrid, Cantoblanco, 28049 Madrid, Spain}
 \author{K. L. Narasimha-Acharya}
   \affiliation{Departamento de F\'{\i}sica de la Materia Condensada, Universidad Aut\'onoma de Madrid, Cantoblanco, 28049 Madrid, Spain}
 \author{F. J. Bailen}
   \affiliation{Departamento de F\'{\i}sica de la Materia Condensada, Universidad Aut\'onoma de Madrid, Cantoblanco, 28049 Madrid, Spain}
 \author{J. J. Palacios}
  \affiliation{Departamento de F\'{\i}sica de la Materia Condensada, Universidad Aut\'onoma de Madrid, Cantoblanco, 28049 Madrid, Spain}
   \affiliation{Instituto de Ciencia de Materiales Nicol\'as Cabrera, Universidad Aut\'onoma de Madrid, Cantoblanco, 28049 Madrid, Spain}
    \affiliation{Condensed Matter Physics Center (IFIMAC),\\ Universidad Aut\'onoma de Madrid, Cantoblanco, 28049 Madrid, Spain}

\date{\today}

\begin{abstract}
We present a theoretical study of the exciton binding energy for anisotropic two-dimensional crystals. We obtain analytical expressions from variational wave functions in different limits of the screening length to exciton size ratio and compare them with numerical solutions, both variational and exact. As an example, we apply these results to phosphorene, a monolayer of black phosphorous. Aided by density functional theory calculations for the evaluation of the two-dimensional polarizability, our analytical solution for the exciton binding energy gives a result which compares well with numerical ones and, in turn, with experimental values, as recently reported.
\end{abstract}

\maketitle

\section{Introduction}

Since the mechanical exfoliation of graphene\cite{Novoselov:science:04}, research in understanding the properties of two-dimensional (2D) crystals has increased in many folds. Atomically thin single-layered materials obtained from transition metal dichalcogenides\cite{Coleman11}, boron nitride\cite{Kawaguchi20081171}, bismuth\cite{PhysRevLett.110.176802}, etc. are being extensively studied for applications as electronic and photoelectronic devices. Few-layered black phosphorous (BP) is a recent addition to the list of graphene-inspired materials\cite{Li14,Liu14,Tran14,2053-1583-1-2-025001}. Apart from having a sizeable band gap which can be tuned by the manipulation of the number of layers, the atomic structure of BP is highly anisotropic which leads to high asymmetry of the electronic band structure even for few layers. In particular, a single layer of BP or phosphorene is attracting most of the attention. The peculiar anisotropic nature of the band gap distinguishes  phosphorene from other 2D crystals, increasing its potential functionality. 

Excitons are a bound state of an electron and an hole and play an important role in the optical properties of the material. Understanding the nature of excitons and their dependence on the electronic structure of the host material is critical and lends a deeper perspective into the many-body physics involved in 2D crystals. The 2D nature of the polarizability of these crystals introduces an important length scale (screening length) $r_0$. For distances between charges in the crystal plane, $r$, greater than $r_0$ the electron-hole binding potential behaves like in a 3D system i.e., it goes as $\sim 1/r$.  However, for the case where $r$ is less than $r_0$ the potential is 2D-like, i.e., logarithmic. This behavior makes excitons in 2D crystals different from their 3D counterparts\cite{PhysRevB.84.085406,PhysRevB.88.045318}. 

Since most common 2D crystals are isotropic, the effect of anisotropy on the optical properties of these materials has remained essentially unexplored. The appearance of phosphorene has, however, changed this view and recent works address this issue from an analytical\cite{2053-1583-1-2-025001}, numerical\cite{Rodin14}, and first-principles\cite{Tran14} standpoints.
Here we give a detailed account of a variational approach, introduced by us in Ref. \onlinecite{2053-1583-1-2-025001}, to the calculation of the  exciton binding energy $E_X$ in anisotropic 2D crystals. Several analytical expressions are derived in certain limits of the 2D interaction potential. The accuracy of our analytical expressions for $E_X$ is tested against both variational and exact numerical solutions to the actual 2D potential, finding excellent agreement in a wide and experimentally relevant range of screening lengths. In particular, the value of $E_X$ for phosphorene, as obtained from our analytical expression, compares almost exactly to the numerical results. Furthermore, this value nicely agrees with the recently reported experimental result\cite{Wang15}.

The present work is divided as follows. In Sec. \ref{sec1} we review the form and limiting behaviour of the Coulomb interaction potential for charged particles in 2D systems. In Sec. \ref{sec2} we present our variational approach based on an anisotropic exciton wavefunction. We first present the analytical result for $E_X$ in the limiting case where the 2D potential reduces to the standard 3D Coulomb potential $\sim 1/r$  for isotropic 2D systems to later introduce the  anisotropy and re-derive the binding energy for this case. In the same manner we derive analytical expressions for the isotropic and anisotropic binding energies in the opposite limit where  the 2D interaction potential behaves logarithmically. We also compare our analytical  expressions with the numerically solved variational problem as well as with the exact numerical solution. In Sec. \ref{sec3} we propose an alternative variational approach based on gaussian orbitals. In Sec. \ref{sec4}, after computing the 2D polarizability with density functional theory (DFT), our analytical approach is applied to the case phosphorene. Finally we present our conclusions in Sec. \ref{sec5}.  

\section{Binding particle-hole potentials  in 2D}
\label{sec1}
As originally derived by Keldysh\cite{Keldysh1979},
the Coulomb potential energy created by a point charge at the origin that electrons feel in 2D layers follows the expression:
\begin{equation}
V_{\rm 2D}(r)=-\frac{e^2}{8\epsilon_0\bar{\epsilon}r_0}\left[
H_0\left(\frac{r}{r_0}\right)- Y_0\left(\frac{r}{r_0}\right)\right],
\label{Keldysh}
\end{equation}
where $r_0\equiv d\epsilon/(\epsilon_1+\epsilon_2)$ and $\bar{\epsilon}=(\epsilon_1+\epsilon_2)/2$. Here $d$ is the thickness of the 2D material, $\epsilon$ is its bulk dielectric constant, and $\epsilon_1$ and $\epsilon_2$ are the dielectric constants of the surrounding media, typically substrate and vacuum. Here $r_0$ plays the role of a screening length and sets the boundary between two different behaviours of the potential. For $r<r_0$ the potential diverges logarithmically, as if created by line charges.  In this limit,  the potential  takes the simplified form also given by Keldysh\cite{Keldysh1979}: 
\begin{equation}
V_{\rm{2D}}(r\ll r_0)\approx \frac{e^2}{4 \pi\epsilon_0\bar{\epsilon}}\frac{1}{r_0}\left[\ln\left(\frac{r}{2 r_0}\right)+\gamma\right],
\label{Log}
\end{equation}
where  $\gamma$ is the Euler constant. For $r>r_0$ the potential becomes the standard Coulomb potential created by point charges which decays as $1/r$: 
\begin{equation}
V_{\rm{2D}}(r\gg r_0)\approx -\frac{e^2}{4 \pi\epsilon_0\bar{\epsilon}}\frac{1}{r}.
\label{Coulomb}
\end{equation}
A very good approximation to the Keldysh potential, fairly accurate in both limits and simpler to use, was introduced by Cudazzo {\em et  al.}\cite{PhysRevB.84.085406}:   
\begin{equation}
V_{\rm{2D}}^{\rm C}(r)= \frac{e^2}{4 \pi\epsilon_0\bar{\epsilon}}\frac{1}{r_0}\left[\ln\left(\frac{r}{r+r_0}\right)+(\gamma-\ln(2))e^{-r/r_0}\right].
\label{Cudazzo}
\end{equation}

\begin{figure}
 \includegraphics[width=\columnwidth]{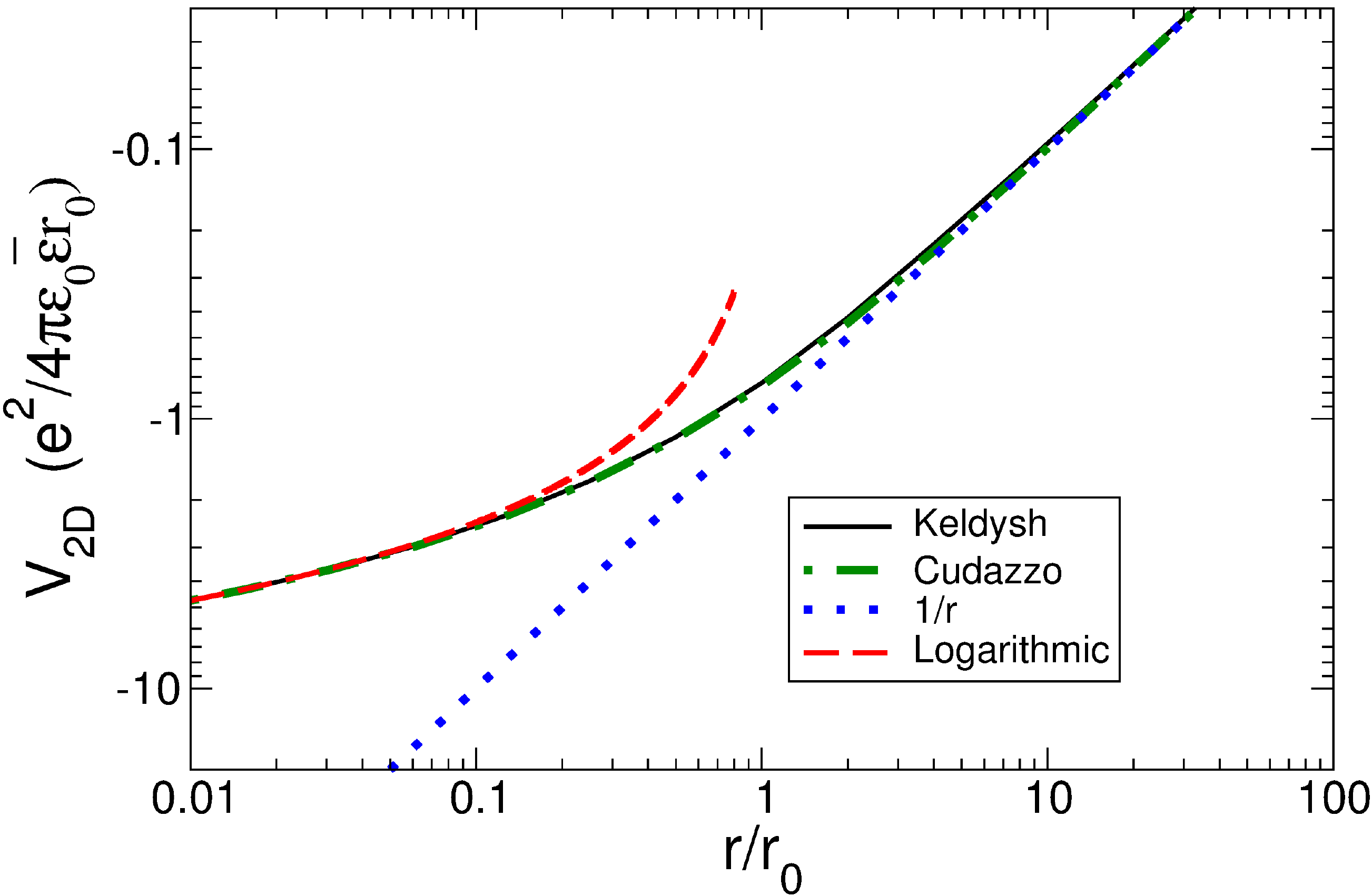}
  \caption{(color online).  Keldysh, 3D Coulomb, logarithmic, and Cudazzo {\em et al.} potentials in log-log scale in a range of distances spanning several decades around $r_0$. }
  \label{Potentials}
\end{figure}

It is interesting to compare these four expressions as a function of the distance $r$ in a range of several orders of magnitude both above and below $r_0$.  We present  such a comparison in Fig.~\ref{Potentials}. There it can seen the range of validity of each approximation, the Cudazzo\textit{ et al.} expression being remarkably accurate for all distances.

\section{variational wavefunction approach}
\label{sec2}
For generic 2D crystals with electrons and holes presenting anisotropic effective masses, $m^{e(h)}_x \neq m^{e(h)}_y$, we consider variational solutions for the exciton wave function of the type\cite{Schindlmayr97}:
\begin{equation}
\phi(x,y)=\left( \frac{2}{a_x^2\lambda \pi} \right)^{1/2} \exp{ \left( - \sqrt{(x/a_x)^2+(y/\lambda a_x)^2} \right) }, 
\label{variational_wavefunction}
\end{equation}
where $\lambda$ is the variational anisotropy scaling factor relating the exciton extension along the $x$ direction, $a_x$ (which is also a variational parameter) and the one along the $y$ direction ($a_y=\lambda a_x$). With this variational wavefunction we can evaluate the expectation value of the kinetic energy:
\begin{align*} 
 E_{\rm kin}(a_x,\lambda) &= \frac{\hbar^2}{2} \iint  \phi \left[ \frac{1}{\mu_x}  \frac{\partial^2 \phi}{\partial x^2}+\frac{1}{\mu_y}\frac{\partial^2 \phi}{\partial y^2}\right] dx dy \\
 &= \frac{\hbar^2}{4a_x^2}\left(\frac{1}{\mu_x}+\frac{1}{\lambda^2\mu_y}\right)
 \label{Kinetic_Energy}
\end{align*}
where $\mu_x$ and $\mu_y$ are the reduced effective masses, $m^e m^h/ (m^e +m^h)$, along $x$ and $y$ directions, respectively.
The  expectation value of the potential energy is given by
\begin{equation} 
 E_{\rm pot}(a_x,\lambda)=\iint V_{\rm{2D}}(x,y)   \phi(x,y)^2 dxdy 
 \label{Potential_Energy}
\end{equation}
and the variational exciton binding energy is obtained from the addition of these two quantities,    
\begin{equation} 
 E_{\rm X}(a_x,\lambda)=E_{\rm kin}+E_{\rm pot}.
  \label{Total_Energy}
\end{equation}
Upon minimization with respect to $a_x$ and $\lambda$, one obtains the optimal parameters defining the extension and shape of the exciton and the actual binding energy $ E_{\rm X}$. Results from three minimization procedures, one analytical and two numerical, are presented in next section.

\subsection{Analytical Results}

 The integral for the potential energy in Eq. (\ref{Potential_Energy}) turns out to be too difficult for an exact variational  analytical solution. The main goal of this section is to make use of the asymptotic behaviour of the Keldysh potential to get analytical expressions for $E_X$ in the limits $r\gg r_0$ and   $r\ll r_0$, namely, valid for large and small excitons, respectively. 

\subsubsection{$r\gg r_0$ limit}

We begin by evaluating  $E_X$ in the isotropic case ($\lambda =1$, $a_x=a_y=a$), considering only the long-range behaviour of the Keldysh potential (see Eq. \ref{Coulomb}). The contribution of the potential energy to $E_X$ is given in this limit by
\begin{equation} 
 E_{\rm pot}= -\frac{e^2}{4\pi\epsilon_0\bar{\epsilon}}\frac{2}{a}.
\end{equation}
Now minimizing $E_X(a)$ with respect to the variational exciton radius one obtains  
\begin{equation}
E_X=-\frac{e^2}{4\pi\epsilon_0\bar{\epsilon}}\frac{1}{\tilde{a}},
\label{EX_3D}
\end{equation}
where the minimal exciton radius $\tilde{a}$ is given by
\begin{equation}
\tilde{a}=\frac{a_0\bar{\epsilon}m}{2\mu},
\label{radius}
\end{equation}
and $m$ and $a_0=\frac{4\pi\epsilon_0}{e^2}\frac{\hbar^2}{m}$ are the free electron mass and the Bohr radius, respectively.

For the anisotropic case ($\lambda \ne 1$)
the exciton extension along the $x$-direction is now given by
\begin{equation}
\tilde{a}_x(\lambda)=\frac{a_0\bar\epsilon m}{4}\left(\frac{1}{\mu_x}+\frac{1}{\lambda^2\mu_y}\right)\frac{1}{I(\lambda)}.
\end{equation}
In the previous expression we find a function of $\lambda$ defined through the elliptic integral
\begin{eqnarray}
&I(\lambda)&\equiv\frac{1}{2 \pi}\int_0^{2\pi}d\theta\frac{1}{\sqrt{1+(\lambda^2-1)\cos^2\theta}}\nonumber\\
&=&\frac{1}{\pi}\left(\frac{K(1-1/\lambda^2)}{\lambda}+K(1-\lambda^2)\right),
\end{eqnarray}
where the function K is the complete elliptic integral of the first kind. Defining now $\mu_{xy}$ as 
\begin{equation}
\mu_{xy}(\lambda)\equiv 2\left(\frac{1}{\mu_x}+\frac{1}{\lambda^2\mu_y}\right)^{-1}I(\lambda),
\end{equation}
the exciton extension along the $x$ axis can now be written as
\begin{equation}
\tilde{a}_x(\lambda)=\frac{a_0}{2}\frac{\bar{\epsilon}m}{\mu_{xy}(\lambda)}.
\end{equation}
The exciton extension along the $y$ direction is thus
\begin{equation}
\tilde{a}_y(\lambda)=\frac{a_0\bar{\epsilon}m}{4}\left(\frac{1}{\mu_x}+\frac{1}{\lambda^2\mu_y}\right)\frac{\lambda}{I(\lambda)}=\frac{a_0}{2}\frac{\bar{\epsilon}m\,\lambda}{\mu_{xy}(\lambda)}
\end{equation}
and the  $\lambda$-dependent 
binding energy of the exciton now becomes 
\begin{eqnarray}
E_X(\lambda)=-\frac{e^2}{4\pi\epsilon_0\bar{\epsilon}}\frac{I(\lambda)}{\tilde{a}_x(\lambda)}
\label{Exciton_Coulomb}.
\end{eqnarray}
We now define
\begin{eqnarray}
&I_E(\lambda)&\equiv(\lambda^2-1)dI(\lambda)/d\lambda+\lambda I(\lambda)\nonumber\\
&=&\frac{1}{\pi}\left(\rm{E}(1-1/\lambda^2)+\frac{\rm{E}(1-\lambda^2)}{\lambda}\right),
\end{eqnarray}
where E is the complete elliptic integral of second kind. We can see that the minimal $\lambda$, $\tilde{\lambda}$, satisfies in general the following equation: 
\begin{equation}
\frac{\mu_x}{\mu_y}=\lambda^3\frac{I_E(\lambda)-\lambda I(\lambda)}{I(\lambda)-\lambda I_E(\lambda)},
\end{equation}
which has no analytical solution for $\lambda$. However, it can be shown \cite{Schindlmayr97} that for $\lambda\lesssim 1$
\begin{equation}
\tilde{\lambda}\approx \left(\frac{\mu_x}{\mu_y}\right)^{1/3}.
\end{equation}
Finally, notice that the results obtained in this subsection will be valid as long as the exciton extension in both $x$ and $y$ directions is much larger than $r_0$. The consistency of this approximation for given experimental parameters ($r_0$, $\mu_x$, $\mu_y$, and $\bar{\epsilon}$) has  to be checked a posteriori. 

\subsubsection{$r\ll r_0$ limit}

As $r \rightarrow 0$ the logarithmic behaviour of the Keldysh potential dominates. The potential energy in 2D takes now the form given in Eq. (\ref{Log}). In the isotropic case the exciton radius is now given by
\begin{equation}
\tilde{a}=\sqrt{\frac{\bar{\epsilon}m}{\mu}a_0r_0}
\label{excitonsize}
\end{equation}
and the binding energy of the exciton is
\begin{eqnarray}
E_X=\frac{e^2}{4\pi\epsilon_0\bar{\epsilon}}\frac{1}{r_0}\left[\frac{3}{2}+\ln\left(\frac{\tilde{a}}{4 r_0}\right)\right].
\label{Eb_log}
\end{eqnarray}

For an anisotropic system the $\lambda$-dependent exciton extension along the $x$ direction is  given by
\begin{equation}
\tilde{a}_x(\lambda)=\sqrt{a_0 r_0\frac{\bar{\epsilon}m}{2}\left(\frac{1}{\mu_x}+\frac{1}{\lambda^2\mu_y}\right)}.
\label{ax}
\end{equation}
Using now a different definition for $\mu_{xy}$
\begin{equation}
\mu_{xy}(\lambda)\equiv 2\left(\frac{1}{\mu_x}+\frac{1}{\lambda^2\mu_y}\right)^{-1},
\label{mu_log}
\end{equation}
the exciton $x$-extension now becomes
\begin{equation}
\tilde{a}_x(\lambda)=\sqrt{a_0 r_0\frac{\bar{\epsilon}m}{\mu_{xy}(\lambda)}}.
\label{excitonsizex_log}
\end{equation}
Again, taking into account that $a_y=\lambda a_x$, the $\lambda$-dependent minimal exciton extension along the $y$ direction is
\begin{equation}
\tilde{a}_y(\lambda)=\sqrt{a_0 r_0\frac{\bar{\epsilon}m\lambda^2}{\mu_{xy}(\lambda)}}.
\label{excitonsizey_log}
\end{equation}
Note that $\tilde{a}_x(\mu_x,\mu_y)=\tilde{a}_y(\mu_y,\mu_x)$.

Finally we obtain the exciton energy for this case:
\begin{equation}
E_X(\lambda)=\frac{e^2}{4\pi\epsilon_0\bar{\epsilon}}\frac{1}{r_0}\left[\frac{3}{2}+\ln\left(\frac{\tilde{a}_x(\lambda)}{4 r_0}\frac{\lambda+1}{2}\right)\right],
\label{EX_log}
\end{equation}
where the minimal $\lambda$ is
\begin{equation}
\tilde{\lambda}= \left(\frac{\mu_x}{\mu_y}\right)^{1/3}
\label{lambda_log}
\end{equation}
for all $\mu_x$ and $\mu_y$.
Again, notice that this result will be valid as long as the $x$ and $y$ minimal extensions of the excitonic wave function are small compared to $r_0$.

Note that Eq. (\ref{EX_log}) can be written in a more symmetrical way as a function of both $\tilde{a}_x$ and $\tilde{a}_y$,
\begin{equation}
E_X(\lambda)=\frac{e^2}{4\pi\epsilon_0\bar{\epsilon}}\frac{1}{r_0}\left[\frac{3}{2}+\ln\left(\frac{\tilde{a}_x(\lambda)+\tilde{a}_y(\lambda)}{8 r_0}\right)\right]\nonumber,
\end{equation}
and that it is also symmetrical under exchange of $\mu_x$ and $\mu_y$, as it should be. However, we find that the binding energy is not only a function of $\mu_x/\mu_y$, but depends on both their values.








Finally, for completeness, we present an analytical expression for the exciton binding energy using the Cudazzo potential in the isotropic case:
\begin{eqnarray}
E_X=\frac{e^2}{4\pi\epsilon_0\bar{\epsilon}}\left[\frac{a_0}{2\mu \tilde{a}^2}+4(\gamma-\ln(2))\frac{r_0}{(\tilde{a}+2 r_0)^2}\right.\nonumber\\
-\frac{1}{r_0}\left(\gamma+\ln\left(\frac{2r_0}{\tilde{a}}\right)\right)\nonumber\\
\left.+\frac{\tilde{a}-2r_0}{\tilde{a}r_0}e^{2r_0/\tilde{a}} Ei\left(\frac{-2r_0}{\tilde{a}}\right)\right],
\label{var_Cudazzo}
\end{eqnarray}
where $Ei$ is the exponential integral function. We have only been able to obtain a working analytical expression for $\tilde{a}$ (too cumbersome to be shown here) in the limit $r_0 \gtrsim a_0$ where the above expression is 
actually useful.

\begin{figure}[t]
\centering{\includegraphics[width=\columnwidth]{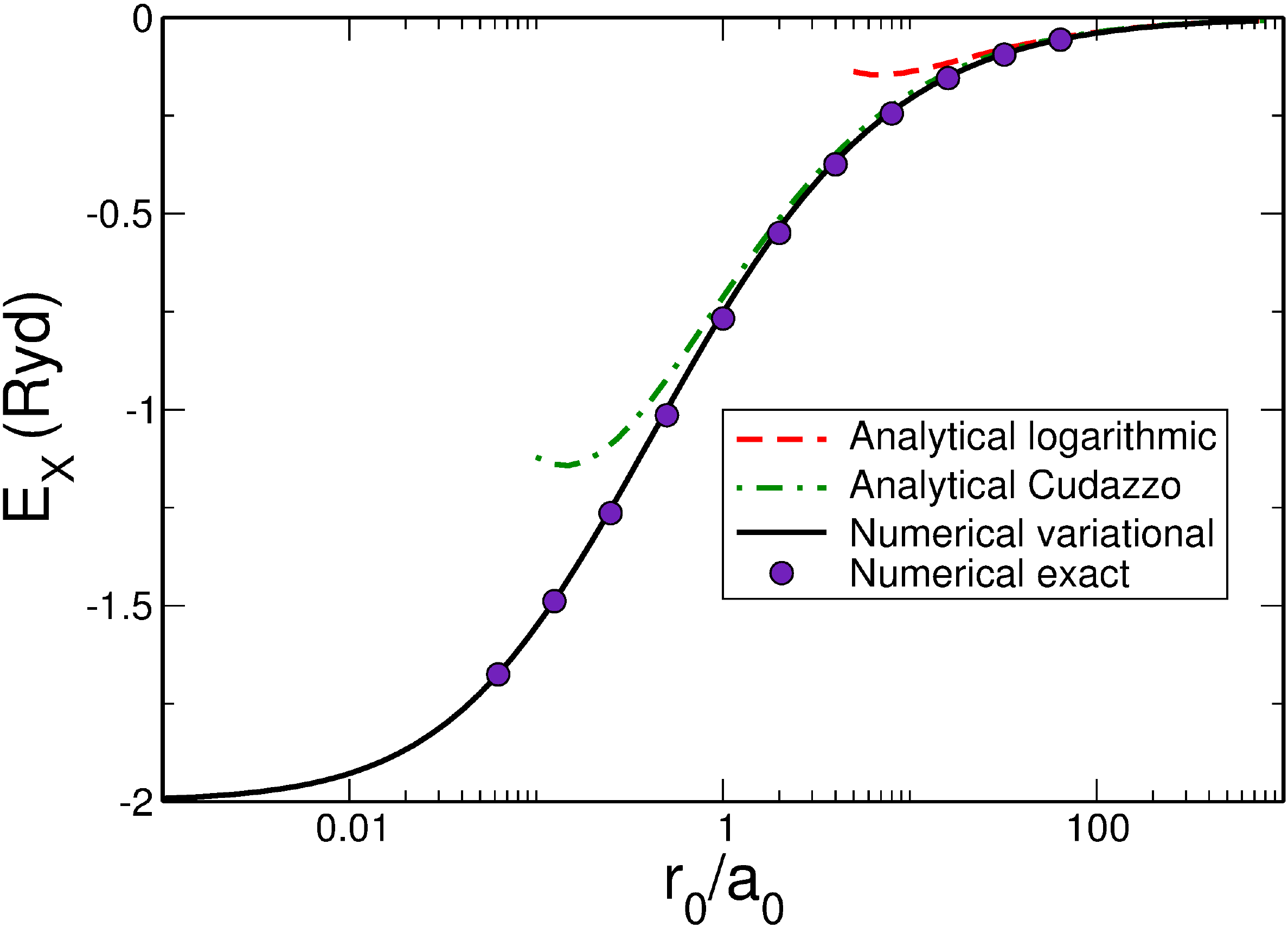}}
\caption{(color online). Binding energies (in Ryd)  for the isotropic exciton computed in different ways: Numerical optimization of the variational wavefunction  in Eq. \ref{variational_wavefunction} (solid line), numerical solution of the  Schr\"odinger equation (circles), analytical variational expression  valid for  large values of  $r_0$  (Eq. \ref{Eb_log})  (dashed line), and expression in Eq.  
\ref{var_Cudazzo}  obtained for the Cudazzo {\it et al.} potential (dotted-dashed line) }
 \label{Isotropic} 
\end{figure} 

\subsection{Numerical Optimization and exact solution}

\begin{figure}[!htbp]
\subfloat[]{\centering{\includegraphics[width=\columnwidth]{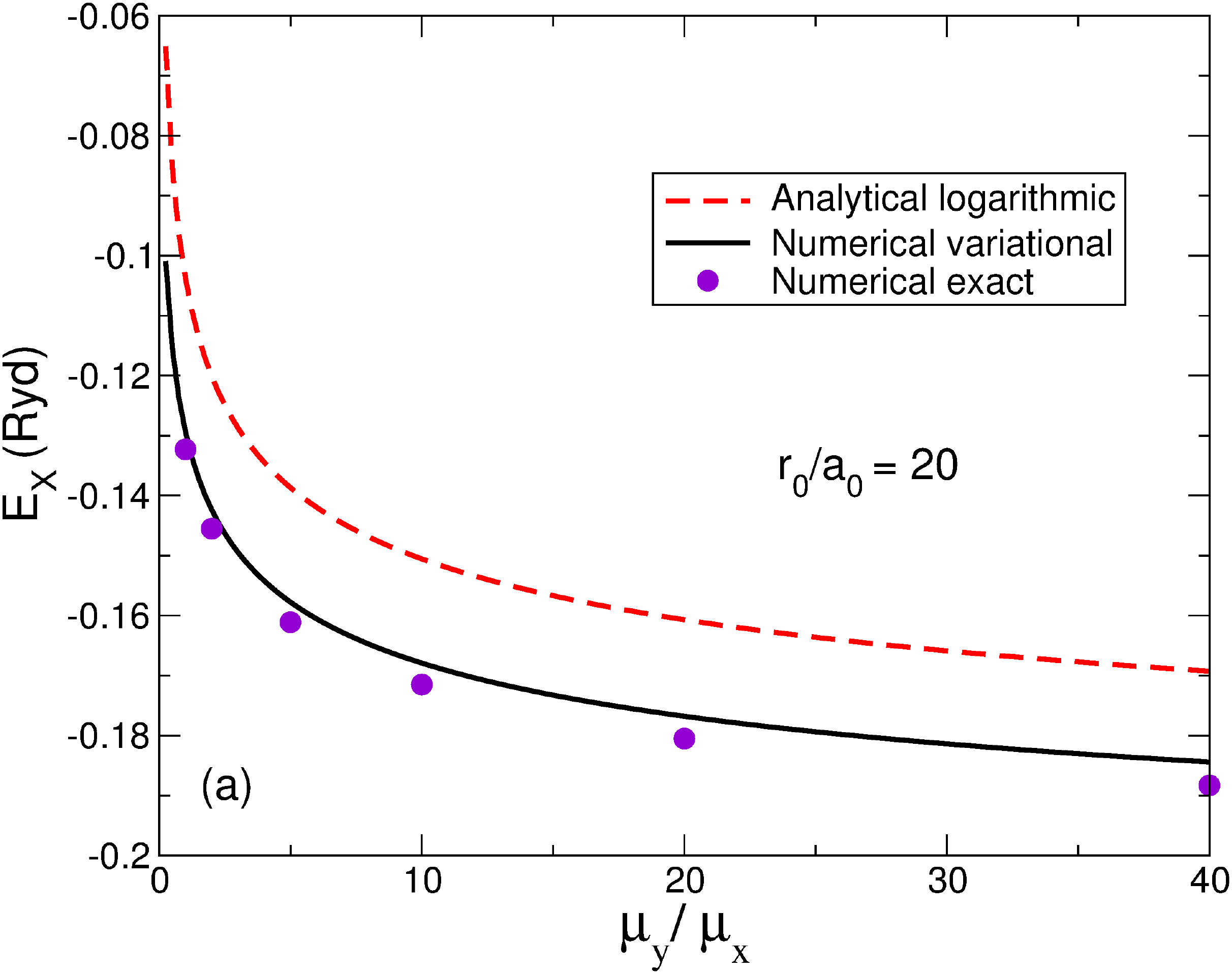}}}\\
\subfloat[]{\centering{\includegraphics[width=\columnwidth]{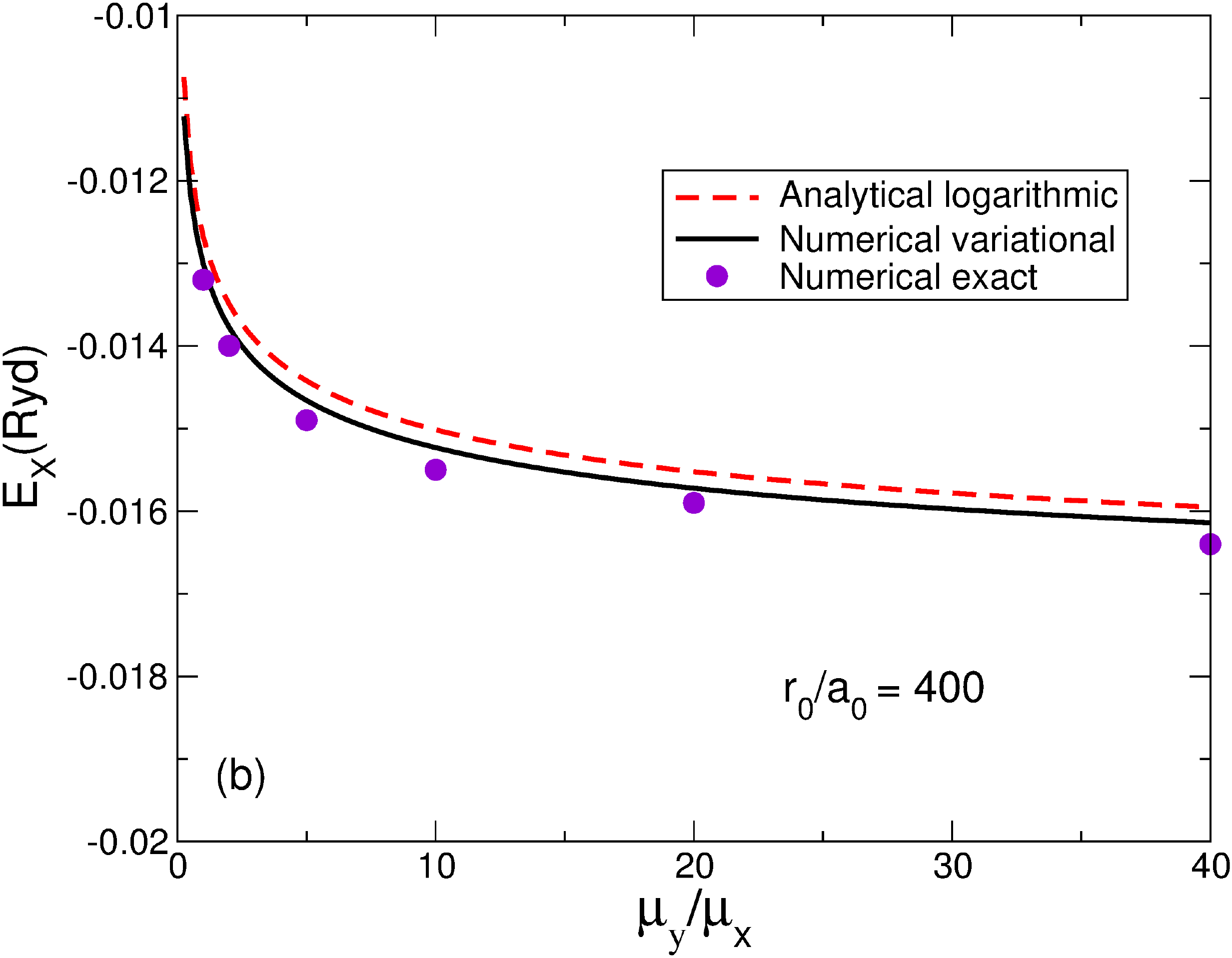}}}
\caption{(color online). Exciton binding energy as a function of the anisotropy $\mu_y/\mu_x$ (for fixed $\mu_x=m/2$) as obtained from the analytical and the numerical approaches for $r_0 = 20 a_0$ (a) and $r_0= 400 a_0$ (b). }
\label{Energy_ans}
\end{figure}  

To validate and test the accuracy of the limiting analytical expressions given in the previous section, we now use the wavefunction in Eq. (\ref{variational_wavefunction}) to numerically compute the potential energy given by Eq. (\ref{Potential_Energy}) for the exact Keldysh potential. We also solve, numerically as well, the 2D Schr\"odinger equation for the same potential, which will give us the exact value of $E_X$ (down to the required numerical precision). Exciton
binding energies for  the isotropic case ($\lambda=1$) are presented in Fig.~\ref{Isotropic} as a function of the screening length $r_0$. For comparison's sake, we take $\bar{\epsilon}= 1$, i.e., the 2D crystal is suspended in vacuum, and $\mu_x=\mu_y=m/2$. Thus, according to Eq. (\ref{EX_3D}), $E_x=-2$ Ryd (where Ryd is the Rydberg energy $13.6$ eV) for $r_0=0$.   The numerical variational result compares  very well  with the exact numerical value in the large range of explored screening lengths. For $r_0 \gg a_0$ the analytical solution in Eq. (\ref{Eb_log}) works fairly well. There, the size of the exciton is smaller than $r_0$ and the  $1/r$ contribution to the Keldysh potential is negligible. As expected, the analytical solution starts to fail as $r_0 \rightarrow a_0$ since there the size of the exciton becomes comparable to $r_0$ and the long-range $1/r$ contribution to the Keldysh potential becomes dominant. (One should keep in mind that the limit of validity of the analytical result, as shown in Eq. (\ref{excitonsize}), depends on the values of $\bar{\epsilon}$ and $\mu$.) We also compare with the result given by Eq. (\ref{var_Cudazzo}), obtained using the approximate expression to the potential in Eq. (\ref{Cudazzo}). This expression, although not as friendly as the previous one, extends the limit of validity of our analytical results down to $r_0\approx a_0$.

The results for the anisotropic case are presented in Fig.~\ref{Energy_ans} for $r_0=20 a_0$   and $r_0=400 a_0$ as a function of the anisotropy ratio $\frac{\mu_y}{\mu_x}$ with $\mu_x=m/2$ (notice a difference of one order of magnitude in the energy scales of each plot). Note that these curves would be identical if plotted as a function of $\frac{\mu_x}{\mu_y}$ with $\mu_y=m/2$. Once again there is close agreement between the analytical solution [Eq. (\ref{EX_log})], the numerical optimization, and the exact numerical solution for large $r_0$, while for the smaller value, the analytical solution visibly deviates from the other two.

\begin{figure}[!htbp]
\includegraphics[width=\columnwidth]{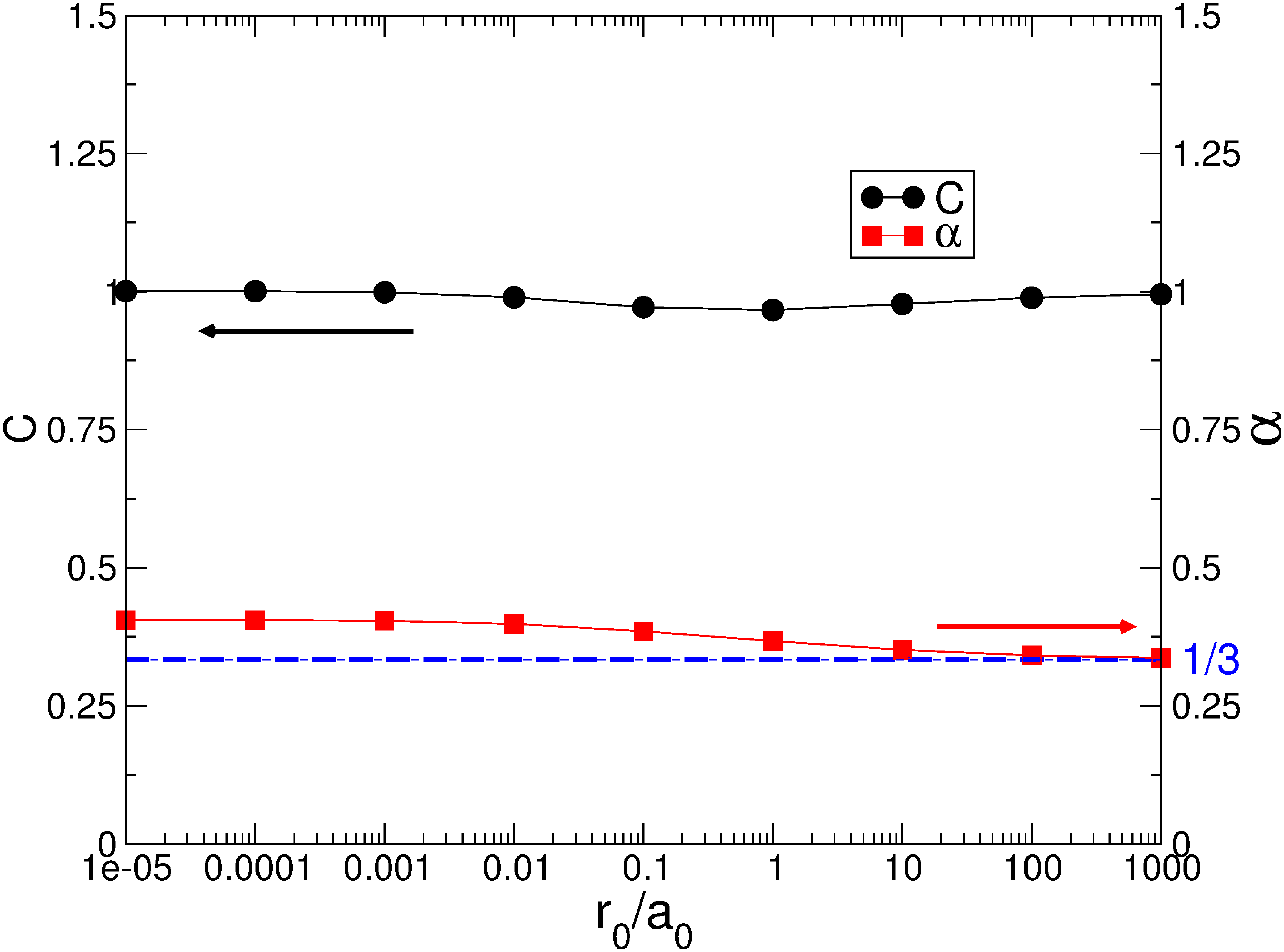}
\caption{ (Color online). Parameters of the fit in expression (\ref{fit_lambda}) as a function of $r_0$:  $C(r_0)$ (left axis) and $\alpha(r_0)$ (right axis).}
 \label{lambda_ans} 
\end{figure}

 An important prediction of our analytical results is the relation between the anisotropy in the exciton extension and the effective masses:    
$\tilde{\lambda}=\left(\frac{a_y}{a_x}\right) \sim \left(\frac{\mu_x}{\mu_y}\right)^{1/3}$, which becomes exact in the limit of small excitons. To test this relation we 
fitted the optimal value of the variational parameter $\tilde{\lambda}$ to the law 
\begin{equation}
\tilde{\lambda}=C(r_0)  \left(\frac{\mu_x}{\mu_y}\right)^{\alpha(r_0)}
\label{fit_lambda}
\end{equation}
 for a large range of $r_0$. The results of this fit are presented in Fig.~\ref{lambda_ans}. 
They  confirm our analytical results and recover the exact 1/3 exponent in the limit of large $r_0$. 

\begin{figure}
\includegraphics[width=\columnwidth]{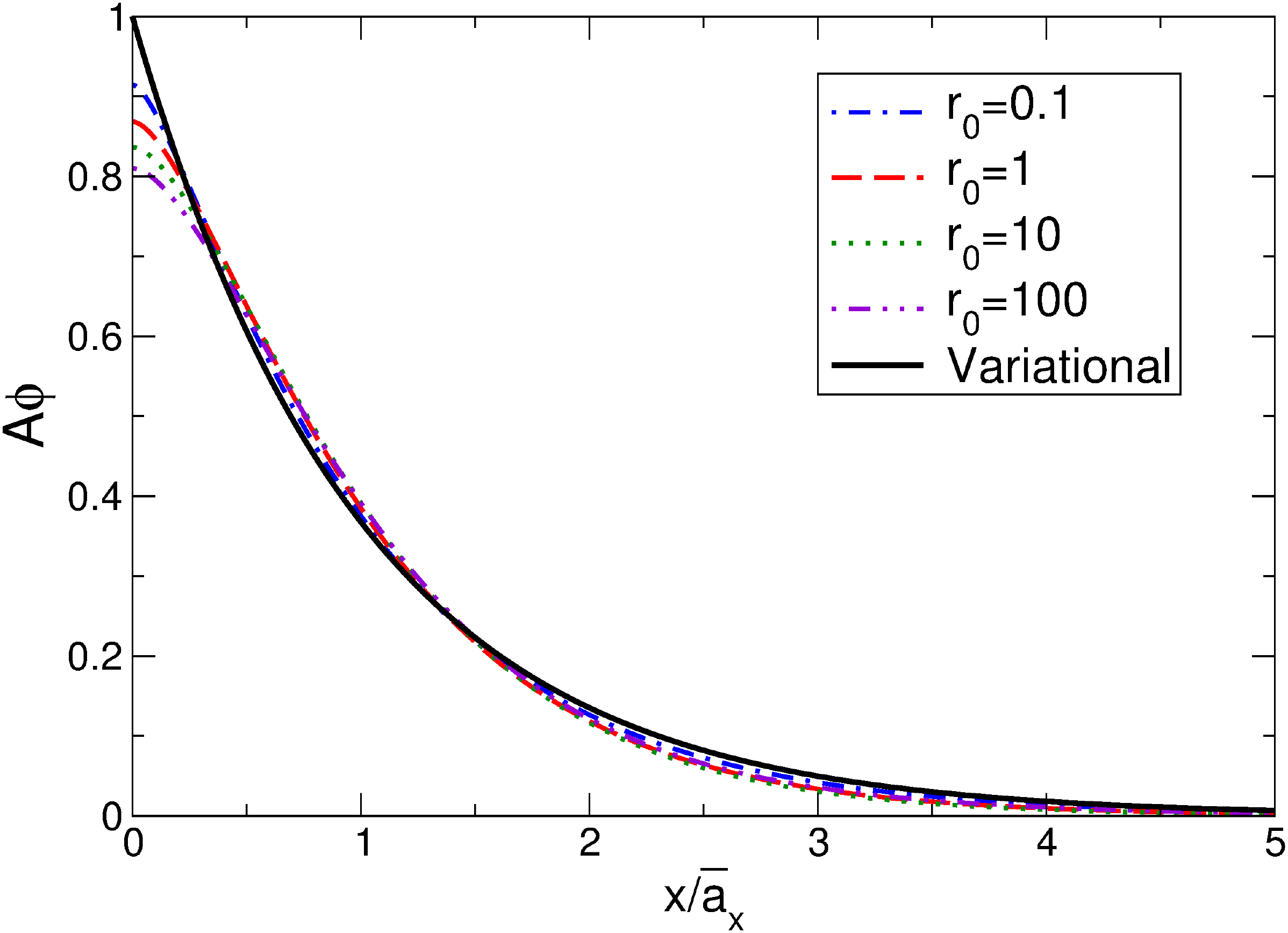}
\caption{(Color online). Exact (dashed lines)  and variational (solid line) wave functions for $r_0=0.1,1,10,100$. 
The x-axis is rescaled with the {\em variational} extension of the exciton and the amplitude 
with the normalization constant of the variational wave function  $A=\left( \frac{2}{a_x^2 \pi} \right)^{1/2}$.
}
 \label{scaling_wf} 
\end{figure}

 We finally provide a comparison between the exact and variational wave functions for several values of $r_0$ in the isotropic limit (see Fig.  \ref{scaling_wf}). Note that the distance is rescaled with the optimal radius and the amplitude of the wave function with the normalization constant  $A=\left( \frac{2}{a_x^2 \pi} \right)^{1/2}$.This representation illustrates to what extent  the exact and variational wave functions satisfy similar scaling relations.  Note that at $r=0$ the exact wave functions do not show the prominent cusp of a 1s Slater-type orbital. This softened behavior at the origin suggests than a combination of gaussian functions may capture more accurately  this feature of the wave function, as shown in next section.

\section{Gaussian-Basis Variational Method }
\label{sec3}
We have found that in the limit of very small $r_0$, the binding energy  is very sensitive to small changes in $r_0$. Furthermore,  the numerical solution of the  Schr\"odinger equation requires a very fine mesh to reproduce the bound state in such a limit. On the other hand, the analytical result found in the $1/r$ limit of the potential constitutes an isolated point an thus cannot be easily extended to small but finite values of $r_0$.  It is therefore interesting to find an alternative numerical method to study  anisotropic excitons in the limit of small $r_0$.    Moreover, as we have presented in Fig. \ref{scaling_wf}, the behavior of the  exciton exact wave functions for different values of the screening length $r_0$ resembles more a 1s Gaussian than a Slater-type orbital.  Gaussian-type orbitals  (GTO) are very efficient basis sets used  intensively in quantum chemistry and solid state calculations. 

The gaussian basis functions $\left\{ \chi_p \right\} $  follow the expression:

 \begin{equation} \chi_p=e^{-\left(\alpha_{x}^{p}x^2+\alpha_{y}^{p}y^2\right)}.
\label{GTO}
\end{equation}

The index $p$ is an integer that here has been chosen to run from 1 to 4 and the exponents $\alpha$ are coefficients that have to be optimized to minimize the ground state energy obtained by the variational method.  Unlike the conventional gaussian  approach, the anisotropy of the problem introduces two coefficients  $\alpha_{x}^{p},\alpha_{y}^{p}$ per GTO. We limit the variational freedom assuming that the anisotropy is identical for the four basis functions:  
\begin{equation} 
\alpha_{y}^{p}=\kappa \alpha_{x}^{p}.
\label{anisot}
\end{equation}
In this equation, $\kappa$ is a constant that does not depend on $p$ so that we reduce the number of exponents that we have to optimize  from eight to four.  The variational wave function in the GTO basis is given by 
\begin{equation} 
\phi_{G}(x,y)= \sum_{p=1}^4 C_p \chi_p.
\label{eq2}
\end{equation}
 For fixed values of $\alpha^p$, the energy is computed by generalized diagnonalization of a 4x4 matrix.  
The matrix elements of the kinetic energy are
\begin{equation} 
H^{\rm kin}_{pq}=\frac{\pi}{\overline{M}_{pq}}  \left(\frac{1}{\mu_x}\frac{\alpha_{x}^{p}\alpha_{x}^{q}}{\alpha_{x}^{p}+\alpha_{x}^{q}}
+\frac{1}{\mu_y} \frac{\alpha_{y}^{p}\alpha_{y}^{q}}{\alpha_{y}^{p}+\alpha_{y}^{q}}      \right).
\label{kin}
\end{equation}
The matrix elements of the potential energy are computed by numerical integration,
\begin{equation} 
H^{\rm pot}_{pq}=\iint V_{\rm{2D}}^K(x,y) \chi_p \chi_q dxdy,
\label{pot}
\end{equation}
and the overlap matrix is
\begin{equation} 
S_{pq}= \frac{\pi}{\overline{M}_{pq}},
\label{overlap}
\end{equation}
where  $\overline{M}_{pq}= {\sqrt{(\alpha_{x}^{p}+\alpha_{x}^{q}) (\alpha_{y}^{p}+\alpha_{y}^{q})} }$.

The binding energy and the optimal wave function are obtained by minimizing numerically the energy with respect to the five variational parameters.   Efficient optimization of the energy requires a careful choice of the initial guess for the values of the exponents $\alpha^p$. In our case,   we choose the optimal values  for a 1s orbital of a ``2D hydrogen atom",  taking $\mu_x=\mu_y=m$ and $r_0$ approaching zero. Once  these optimal exponents are obtained,  $r_0$  is changed slightly and the problem is solved again, using this time the exponents $\alpha$ obtained in the previous step. The procedure continues  until the ground state energy for the desired $r_0$ is reached.  
 
\begin{figure}
  \includegraphics[width=\columnwidth]{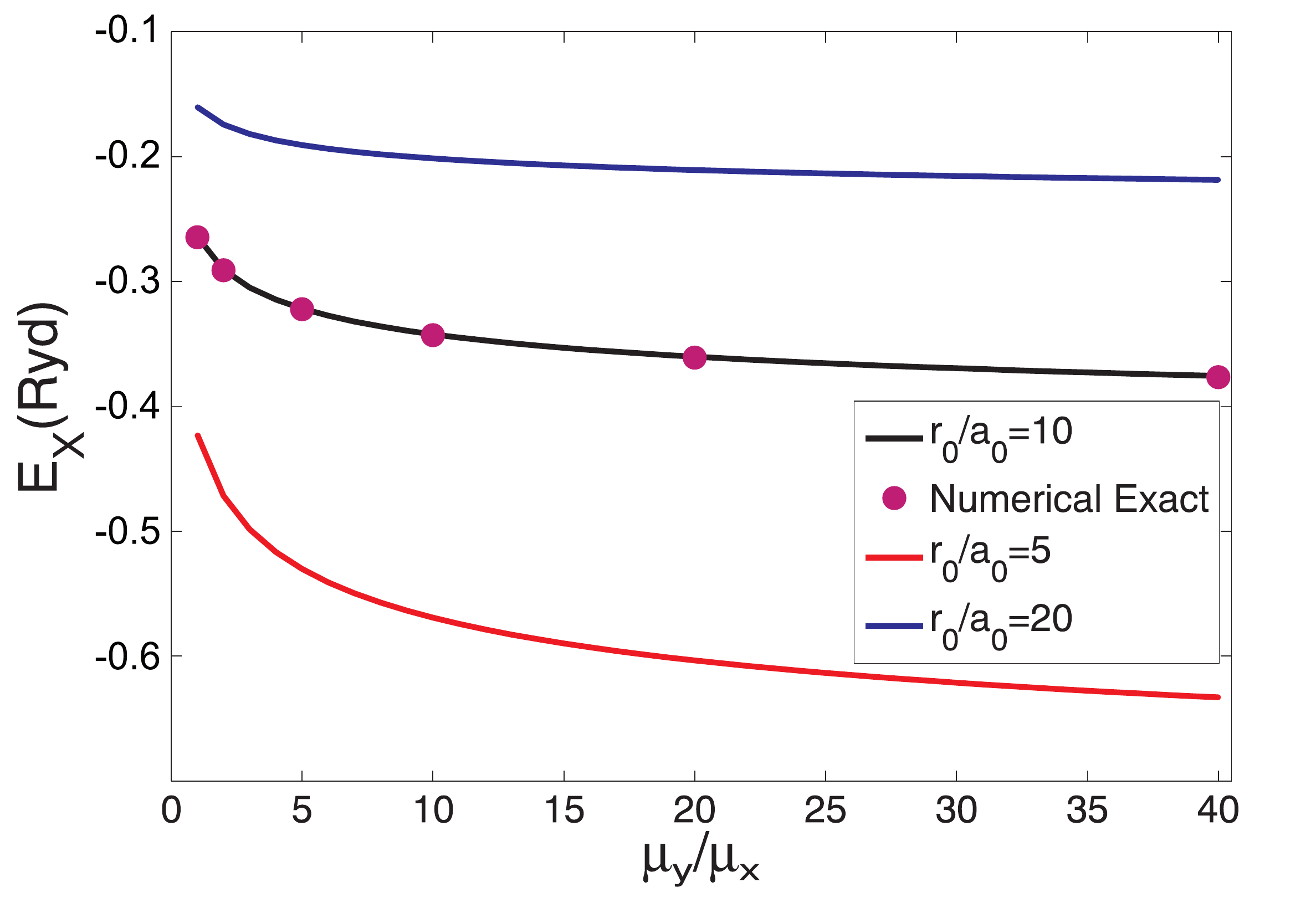}
  \caption{(Color online). Ground state energies obtained using the gaussian variational method (continuous lines) and  the numerical solution of Schr\"odinger equation for $r_0=10 a_0$ (circles) as a function of the  effective mass ratio $\mu_y/\mu_x$ for $\mu_x=m$.}
  \label{Energies_Gaussian}
\end{figure}

For example, fixing $r_0=10 a_0$ and $\mu_x=m$, the effective mass along the $y$ axis $\mu_y$ is varied  from 1 to 40$m$. In this case the initial guess for the exponents is the last set of coefficients $\alpha$ obtained when changing $r_0$. The result for the ground state energy of this calculation is shown in Fig.~\ref{Energies_Gaussian} in comparison to the numerical solution of the Schr\"odinger equation with the Keldysh potential. They match perfectly. Two other values of $r_0/a_0$ obtained with the same gaussian-basis variational method are also shown. 

\begin{figure}
\includegraphics[width=\columnwidth]{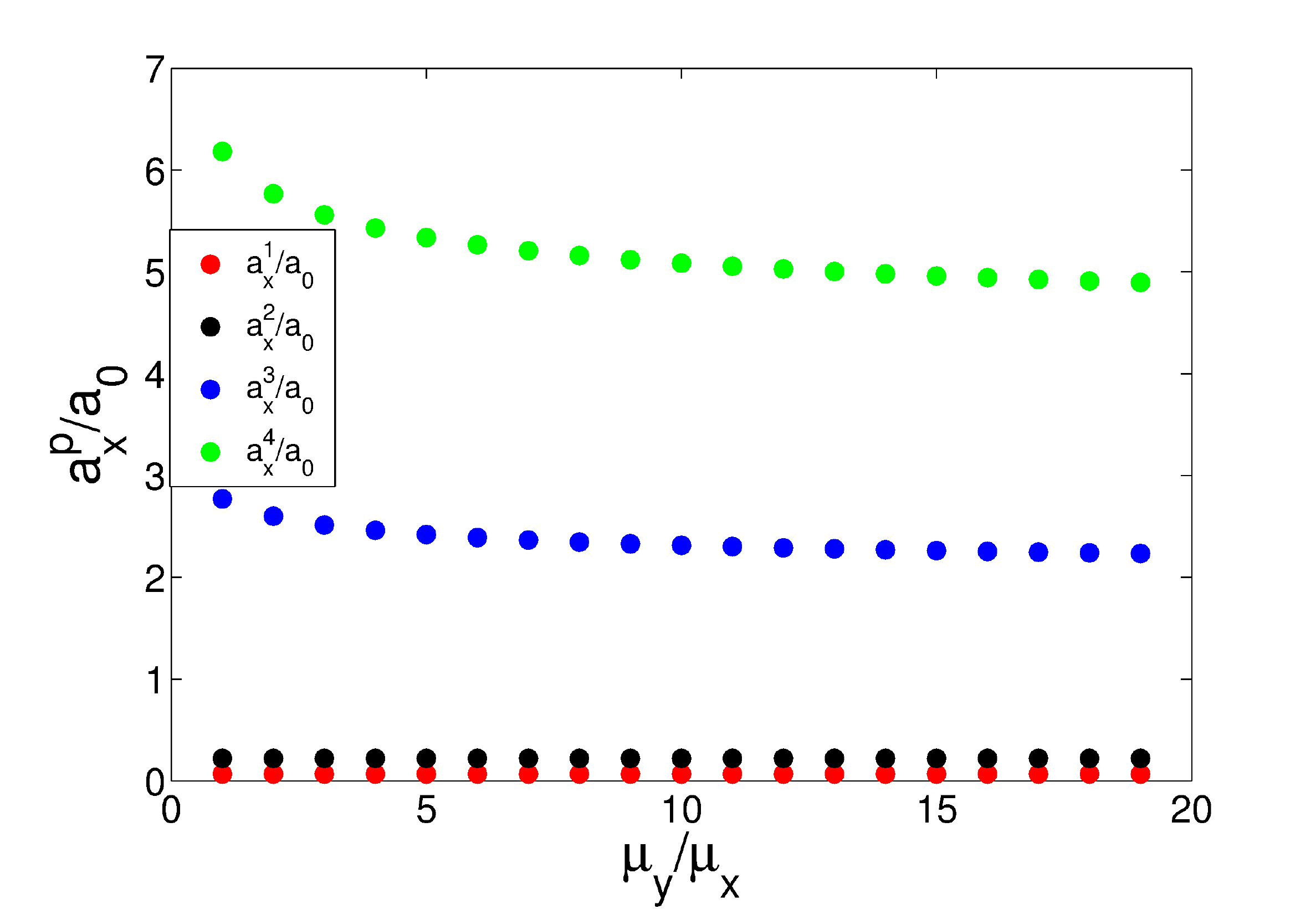}
\caption{(Color online). Coefficients $a_{x}^{p}=1/\sqrt{\alpha_{x}^{p}}$ for $r_0=10 a_0$ obtained versus $\mu_y/\mu_x$ for $\mu_x=m$.}
\label{Length_Gaussian}
\end{figure}

\begin{figure}
    \includegraphics[width=\columnwidth]{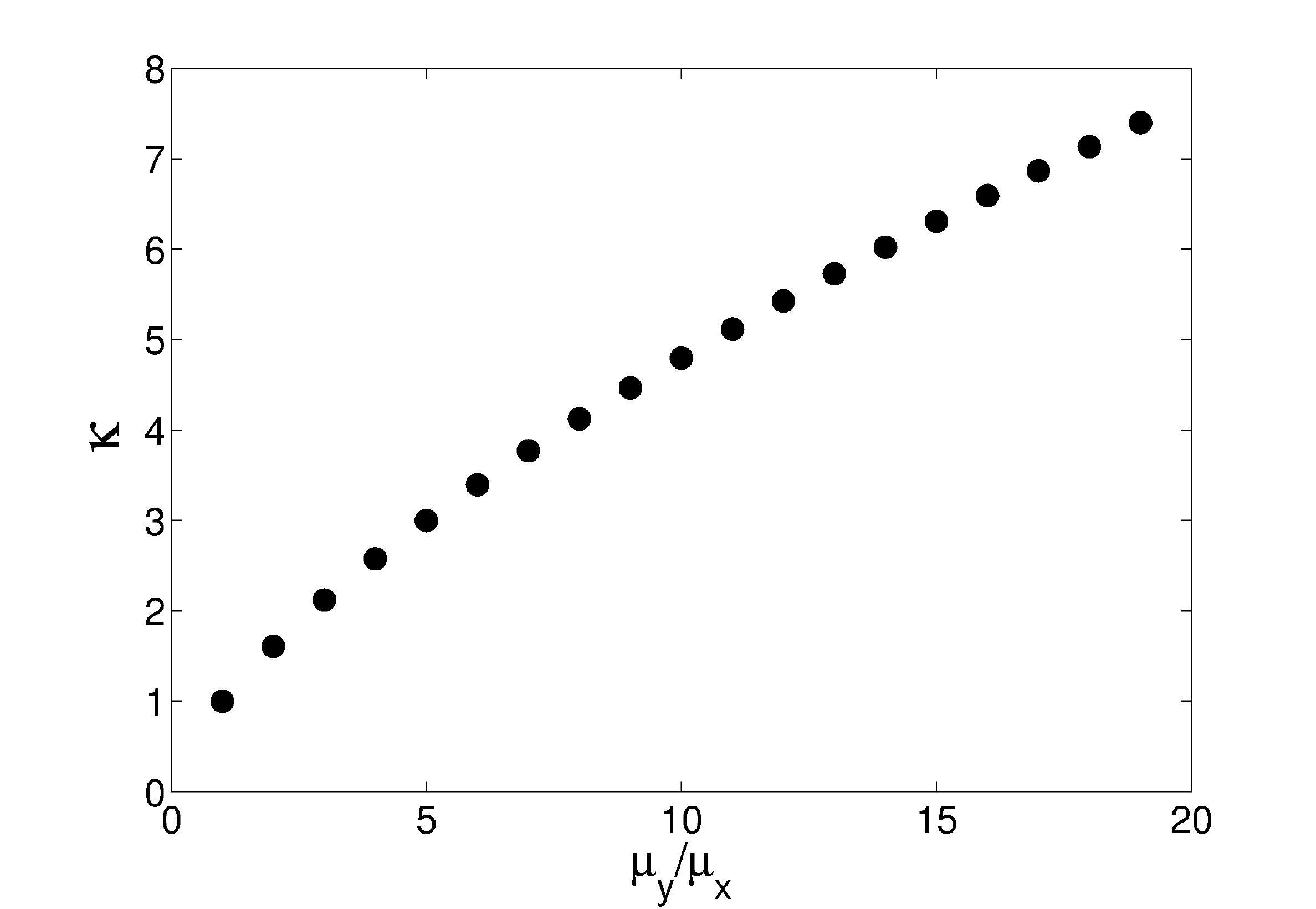}
  \caption{(Color online). Coefficient $\kappa=\alpha_{y}^{p}/\alpha_{x}^{p}$ for $r_0=10 a_0$ obtained versus $\mu_y/\mu_x$ for $\mu_x=m$.}
  \label{fig3}
\end{figure}

\begin{figure}
   \includegraphics[width=\columnwidth]{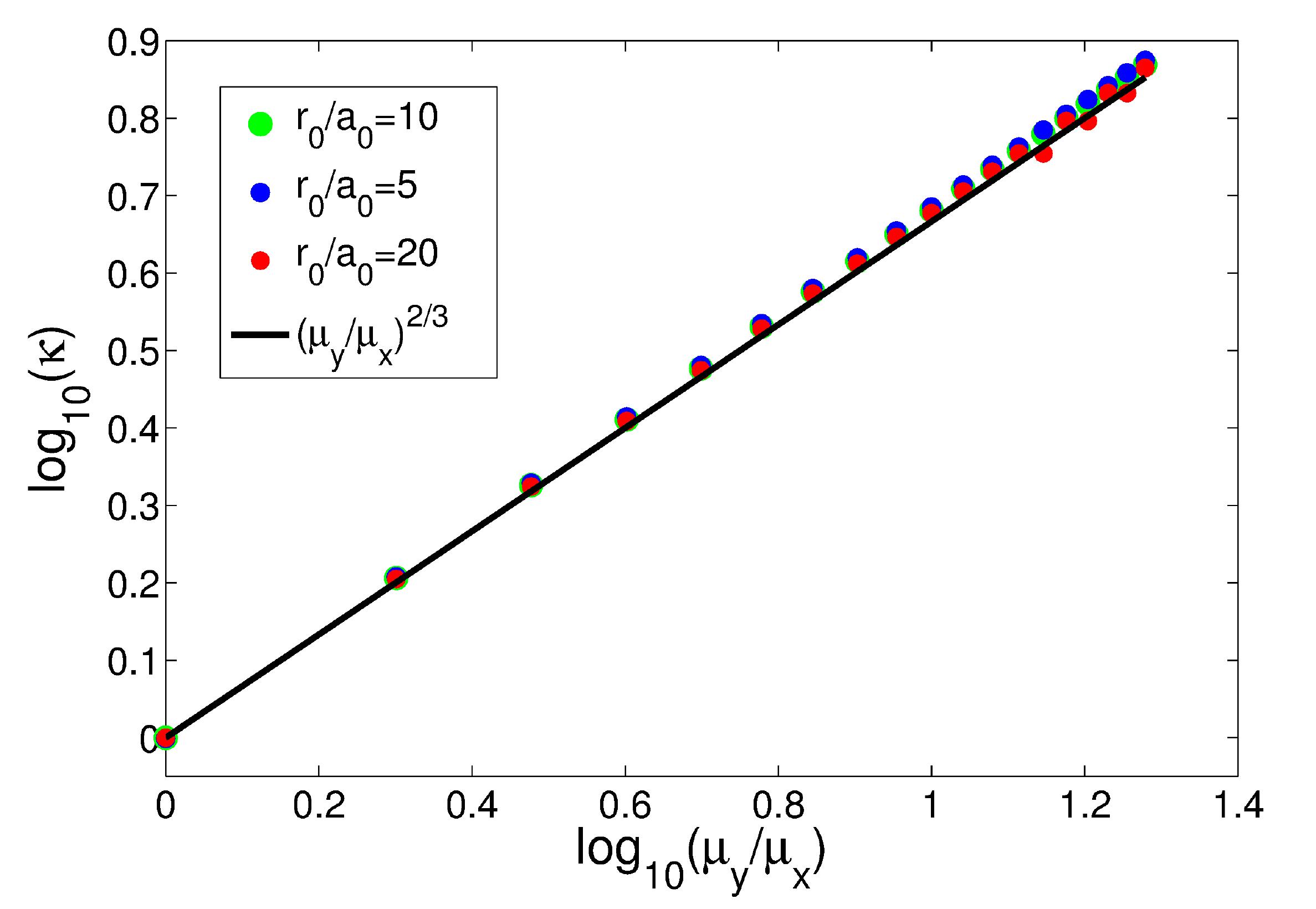}
  \caption{(Color online). Log-log representation of the coefficient $\kappa=\alpha_{y}^{p}/\alpha_{x}^{p}$ obtained for different values of $r_0$ (circles) and its linear fitting (continuous line) versus $\mu_y/\mu_x$ for $\mu_x=m$.  }
  \label{fig4}
  \end{figure}

In Fig.~\ref{Length_Gaussian}, the length scales $a_{x}^{p}={1}/{\sqrt{\alpha_{x}^{p}}}$ of the whole set of optimal GTO's for $r_0=10 a_0$ are plotted vs $\mu_y/\mu_x$ for $\mu_x=m$. 
$\kappa$ is also plotted in Fig.~\ref{fig3} against the same quantity. In Fig.~\ref{fig4} we show a log-log representation of $\kappa$ versus the asymmetry ratio for different values of $r_0$ where a linear fitting has been made.

\section{Application to phosphorene}
\label{sec4}
A paramount example of an anisotropic 2D crystal is phosphorene\citep{Li14,Liu14,2053-1583-1-2-025001}, where the effective masses along $x$ and $y$ directions can differ by even an order of magnitude. From the start we have chosen to express the 2D potential constant in terms of the bulk dielectric constant $\epsilon$ and the effective thickness of the 2D crystal $d$ (see Eq. \ref{Keldysh}).  
Equivalent expressions for the 2D potential, which rely on the evaluation of the actual 2D polarizability of the 2D crystal, $\chi$, have recently been proposed\cite{PhysRevB.84.085406,PhysRevB.88.045318,Rodin14}. These are probably more appropriate for actual crystals, although it has also been shown that Eq. (\ref{Keldysh}) works well as long as $\epsilon$ is taken as the in-plane component of the bulk dielectric tensor\citep{PhysRevB.88.045318} of the 3D crystal. Here we will compare both possibilities.

As shown in Ref. \onlinecite{PhysRevB.84.085406}, the screening length $r_0$ depends on the polarizability $\chi$ as  $r_0\equiv 2\pi\chi$. The polarizability for 2D materials can be computed using the expression
 \begin{equation}
 \epsilon(L)= 1+\frac{4\pi\chi}{L},
 \label{chi}
 \end{equation}
where $L$ is the distance between layers in a 3D layered structure. As can be seen, the dielectric function $\epsilon$ tends to unity as the inter layer distance $L$ tends to infinity. We have computed the dielectric function at different inter layer distances within the density functional theory framework using the Perdew-Burke-Ernzerhof (PBE) functional\cite{Zhang98} and norm conserving Troullier-Martins (TM) pseudopotentials as available in the SIESTA package\cite{SIESTA}. The atomic and electronic structures have been duly converged on all parameters. SIESTA calculates the imaginary part of the dielectric function from which the real part of it is obtained using the Kramers-Kronig relations. In order to account for the under-estimated band gap, the scissor approximation, as implemented in SIESTA, has been utilized. The scissor shift of 1.2505 eV was made to match our previously reported  band gap value of 2.15 eV\cite{2053-1583-1-2-025001}. While more elegant approaches to the gap problem of phosphorene have been reported in the literature\citep{Tran14}, the scissor approximation suffices to our purpose here.

\begin{figure}
 \centering
\includegraphics[width=\columnwidth]{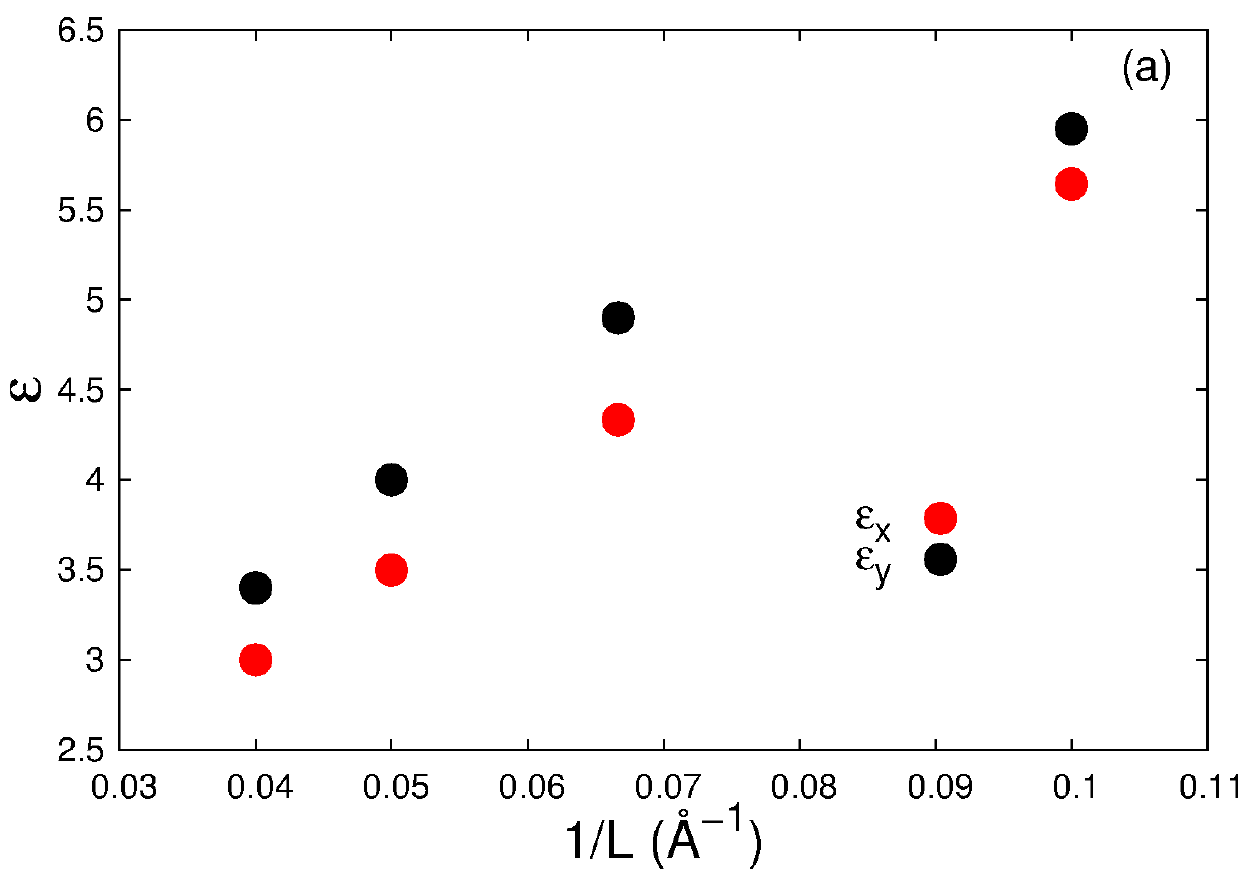}
\includegraphics[width=\columnwidth]{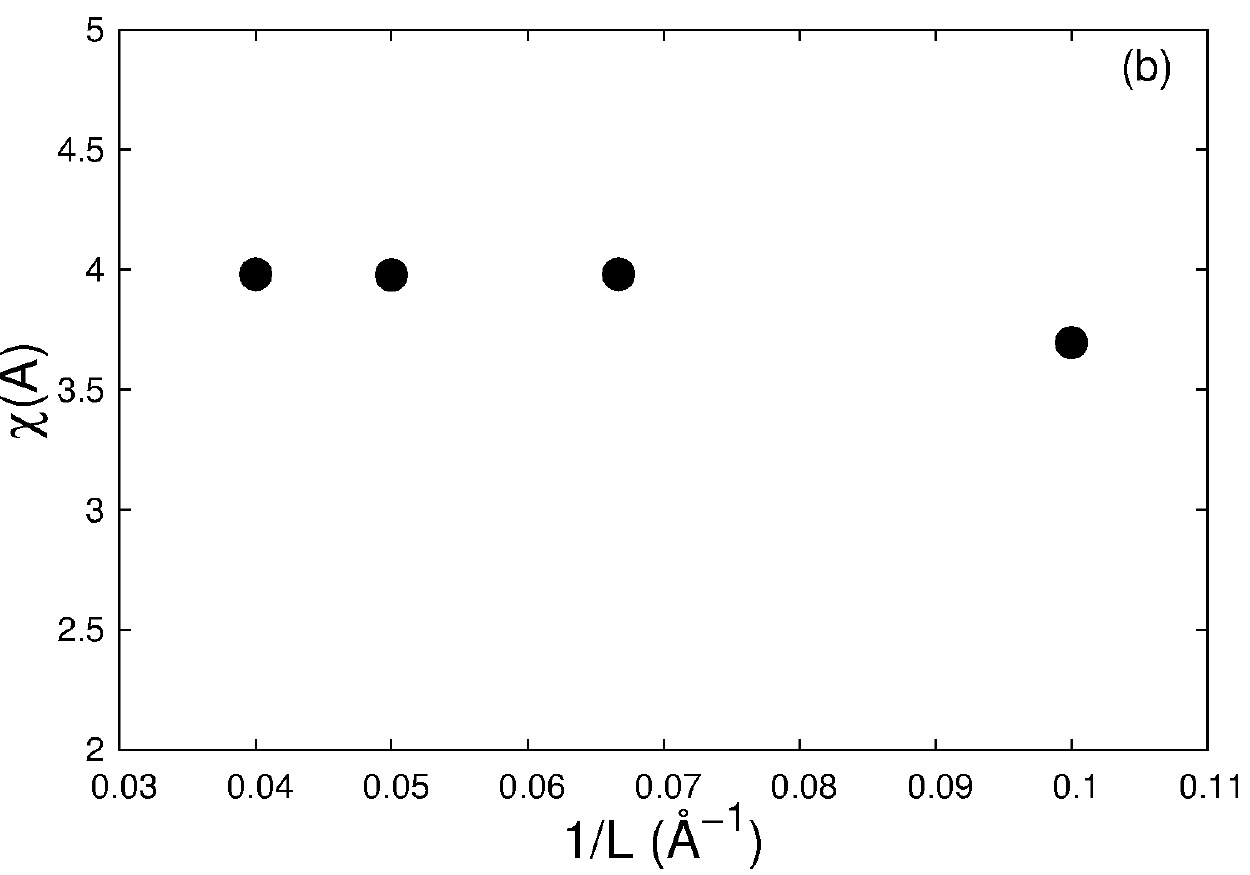} 
 \caption{(Color online). (a) Real part of the $x$ and $y$ components of the dielectric constant (evaluated at zero frequency) for different values of the interlayer distance $L$ in a 3D black phosphorous structure. (b) Polarizability as obtained from Eq. \ref{chi} for different values of $L$ after averaging on the plane.}
\label{fig:dielectric}
\end{figure}
 
Using the plane-averaged static dielectric function calculated with SIESTA (see Fig. \ref{fig:dielectric}), a value in the vicinity of $\chi$ of 3.8 \AA{} is obtained. This value for $\chi$ yields a screening length of $r_0=23.2$ \AA{}. From the numerical variational solution we obtain an exciton extension of $\tilde{a}_x = 11.7$ \AA{} and $\tilde{a}_y =5.9$ \AA{}
along the x and y directions, respectively. Since these values are smaller than $r_0$, the use of the analytical expressions obtained in the logarithmic limit of the potential is justified. Also this was expected from Fig. \ref{Isotropic} and, in particular, from the comparison shown in Fig. \ref{Energy_ans} for anisotropic cases. There it can be seen that already for $r_0=20a_0$ the deviation between the analytical result and the numerical ones is less than 10\% for a ratio $\mu_y/\mu_x \approx 7$ (which corresponds to phosphorene). Using now Eqs. (\ref{ax})–(\ref{lambda_log}), we obtain an exciton binding energy for phosphorene in vacuum of $E_X = 0.61$ eV, while
the numerical variational value is $E_X = 0.78$ eV. This result is remarkably close to a recently reported experimental value of $\approx 0.9$ eV. The agreement is somewhat surprising since this has been measured for phosphorene on a SiO substrate\cite{Wang15}. A more recent experiment, however, reports a smaller value for $E_X$, which is maybe more expected due to the screening of the substrate\cite{Yang15}.

\begin{figure}
 \centering
\includegraphics[width=\columnwidth]{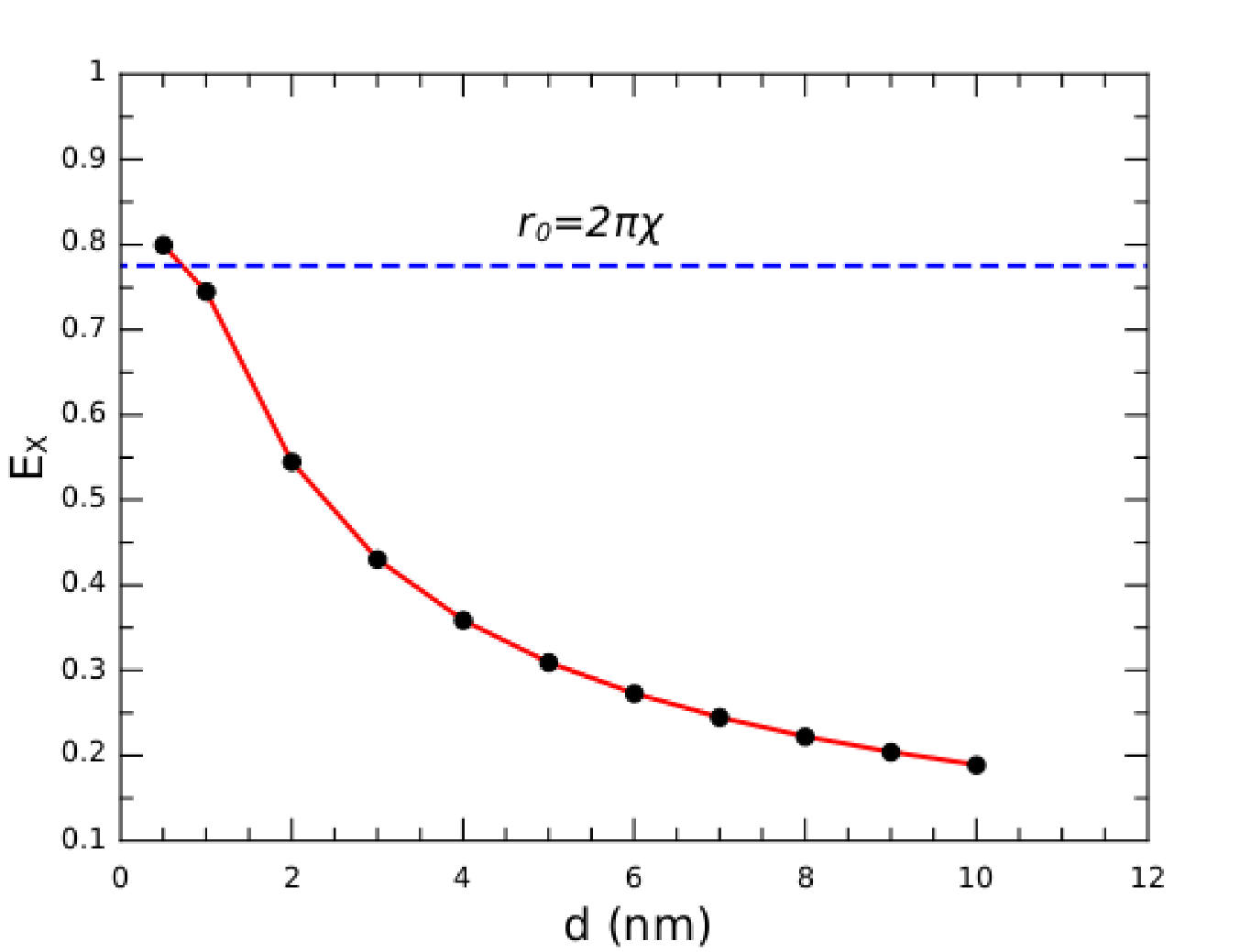}
 \caption{(Color online). Exciton binding energy (in eV) as a function of the  thickness when all the other parameters (effective mass, bulk dielectric constant, and screening length) to correspond to phosphorene. The horizontal line marks the value obtained using the actual 2D polarizability of phosphorene. }
\label{fig:Eb_phos}
\end{figure}

Similarly, we can use the value  of $r_0$ obtained from the real part of the bulk $\epsilon$ (at zero frequency) and the thickness $d$ of the monolayer. Since the thickness is somewhat undetermined, we have computed the binding energy  for varying $d$, as shown in Fig. \ref{fig:Eb_phos}. It can be observed that the binding energy of the monolayer computed using the microscopically derived $\chi$ matches the binding energy obtained using $r_0\equiv d\epsilon/(\epsilon_1+\epsilon_2)$ for $d\approx 7$ \AA{}, which certainly can be considered the thickness of phosphorene.

\section{conclusions}
\label{sec5}
We have shown that a variational approach to the exciton binding energy in anisotropic 2D crystals can give excellent results when compared to numerical approaches. Furthermore, we have studied the range of validity of analytical solutions to the variational approach and found that these can give highly satisfactory results in a range of values of screening lengths which is relevant for actual 2D crystals such as phosphorene. We have computed the exciton binding energy in this case and found a very good agreement with a recently reported experimental result\cite{Wang15}. As long as the screening length to exciton size is large, our analytical results can be trivially used to predict the exciton binding energy of any 2D crystal. 

\acknowledgments{
This work was supported by MINECO under Grants Nos. FIS2013-47328 and FIS2012-37549, by CAM under Grants Nos. S2013/MIT-3007, P2013/MIT-2850, and by Generalitat Valenciana under Grant PROMETEO/2012/011. The authors thankfully acknowledge the computer resources, technical expertise, and assistance provided by the Centro de Computaci\'on Cient\' ifica of the Universidad Aut\'onoma de Madrid.
E.P. also acknowledges the Ram\'on y Cajal Program.}

\end{document}